\documentstyle[pre,aps,epsf,psfig]{revtex}
\newcommand \be[1]{\begin{equation}\label{#1}}
\newcommand \ee{\end{equation}}
\newcommand \beq[1]{\begin{eqnarray}\label{#1}}
\newcommand \eeq{\end{eqnarray}}
\newcommand \bib{\bibitem}

\newcommand{\DK}[1]{\mbox{\boldmath$#1$}}

\newcommand \al{\alpha}
\newcommand \nn{\nonumber}
\newcommand \om{\omega}
\newcommand \eps{\epsilon}
\newcommand \veps{\varepsilon}
\newcommand \arsinh{{\rm arsinh}}
\begin{document}

\draft
\title{
The use of relativistic action in strong-field nonlinear photoionization 
}

\author{J. Ortner $\,^{a)}$ and V.~M.~Rylyuk $\,^{b)}$}
\address{
$\,^{a)}$ {\it Institut f\"ur Physik,Humboldt Universit\"at zu Berlin, 
Invalidenstr. 110, 10115 Berlin, Germany}\\
$^{b)}${\it Department of Theoretical Physics, University of Odessa, 
Dvorjanskaja 2, 270100, Odessa, Ukraine}}
\date{submitted to Phys. Rev. A}
\maketitle
\begin{abstract}
Nonlinear relativistic ionization phenomena induced by a strong linearly polarized laser field are considered. The starting point is the classical relativistic action for a free electron moving in the electromagnetic field created by a strong laser beam. This action has been used to calculate semiclassical transition rates. Simple analytical expressions for the ionization rate, the photoelectron emission velocity  and for the drift momentum distribution of the photoelectron have been found.  The analytical formulas apply to nonrelativistic bound systems as well as to initial states with an energy corresponding to the upper boundary of the lower continuum and to  the tunnel as well as the multiphoton regime. In the case of a nonrelativistic bound system we recover the Keldysh formula for the ionization rate. Relativistic effects in the initial state lead to a weak enhancement of the rate of sub-barrier ionization and to the appearance of a nonzero photoelectron leaving velocity. 
\end{abstract}


\section{Introduction}

Relativistic ionization phenomena induced by strong laser light has become a topic of current interest \cite{CR94,PMK97,PMK98,DK98,K98,CR98,K99}. In the nonrelativistic theory it is assumed that the electron velocity in the initial bound state as well as in the final state is small compared with the speed of light. However, the electrons may be accelerated up to relativistic velocities in an intense electromagnetic field produced by modern laser devices. If the ponderomotive energy of the electron is of the order of the rest energy a relativistic consideration is required. Relativistic effects in the final states become important for an infrared laser at intensities of some $10^{16} {\rm W}\,{\rm cm}^{-2}$.  The minimal intensity required for relativistic effects increases by two orders of magnitude for wavelength corresponding to visible light. The main relativistic effects in the final state are \cite{CR94,DK98,K98,CR98,K99}: (i) the relativistic energy distribution and (ii) the shift of the angular distribution of the emitted electrons towards the direction of propagation of incident laser beam. Relativistic effects have also to be taken into account if the binding energy $E_b$ in the initial state is comparable with the electron rest energy \cite{PMK97}. A relativistic formulation is necessary for the ionization  of heavy atoms or singly or multiply charged ions from the inner shells. 

This paper is aimed to consider the relativistic effects connected with  relativistic final states velocities and/or low lying initial states  from a unique point of view.

Let us start with the classical relativistic action for an electron of charge $e$ moving in the field of an  electromagnetic plane wave with the vector potential $\DK{A}(t-x/c)$. Here and below $\DK{A}$ denotes a two-dimensional vector in the y-z plane. The action may be found as a solution of the Hamilton-Jacobi equation and reads \cite{Landau}
\be{Srel}
S_f(\xi;\xi_0)= mc^2\Biggl\{ \DK{f} \cdot \frac{\DK{r} }{c}-\alpha \frac{x}{c}-\frac{1+\al^2+f^2}{2 \al} \left(\xi-\xi_0\right)+ \frac{e}{m c^2 \al} \DK{f} \int_{\xi_0}^{\xi} \DK{A} d\xi - \frac{e^2}{2 m^2c^4 \al} \int_{\xi_0}^{\xi} \DK{A}^2 d\xi \Biggr\} \,,
\ee
where $\al$ and $\DK{f}=(a_1,a_2)$ are constants, $\DK{r}=(y,z)$; further is $\xi=t-{x}/{c}$, $\xi_0$ is the initial value. By applying the usual Hamilton-Jacobi method we take the derivative of the action  $S_f$ with respect to the constants  $a_1,a_2$ and $\alpha$ and set the result equal to new constants $\beta_1,\beta_2$ and $\beta_3$ in order to obtain the electron trajectory under the influence of the wave field. Assuming a harmonic plane wave of linear polarization with the electric field $\DK{E}=F {\DK{e}_y\cos{\om\xi}}$ we obtain that the electron motion in the field and in the laboratory coordinate system is given by
\beq{trajec}
\alpha^2 (t+x/c) - \beta^2 \xi+\frac{2 \eps}{ \om}
a_1\cos{\om\xi}
+\frac{1}{4 \om} \eps^2 \sin{2\om\xi}= \beta_3&,&~~~
v_x = c \frac{f(\xi) -1}{f(\xi) +1} \,,\nn\\
y = \beta_1 +\frac{ca_1}{\alpha} \xi -\frac{c \eps}{ \alpha\om}
\cos{\om\xi}\,&,&~~~
v_y =\frac{2c}{ \alpha (1+f(\xi))} \left\{a_1 +
\eps  \sin \om \xi \right\},\nn \\
z =\beta_2 +\frac{ca_2}{\alpha} \xi \,&,&~~~
v_z =\frac{2c}{ \alpha (1+f(\xi))} a_2 \,,\nn\\
f(\xi)= \frac{\delta ^2}{\alpha^2} +
\frac{2\eps}{\alpha^2} a_1\sin{\om\xi}+ \frac{1}{\alpha^2} \eps^2
\sin^2{\om\xi}&,&
\eeq
where $\beta_1,\beta_2$ and $\beta_3$ together with $a_1,a_2$ and $\alpha$ have to be determined from the initial conditions for position and velocity. Further we have introduced the notations $\beta^2 =1+a_1^2+a_2^2 +\eps^2/2$,  $ \delta ^2 =1+a^2_1 + a^2_2$ and the parameter $\eps=e F/\om m c $ characterizing the strength of relativistic effects.

Consider now the process of nonlinear ionization of a strongly bound electron with a binding energy $E_b$ comparable with the rest energy. Recently  the ionization process in static crossed electric and magnetic fields has been considered \cite{PMK97,PMK98}. The results of this paper may be applied to the ionization in laser fields only for the case of very strong fields $\eps \gg 1$. With an increasing frequency of the laser light (especially for a tentative x-ray laser) very high laser intensities are required to satisfy this condition. Therefore it is necessary to generalize the result of \cite{PMK97,PMK98} to the case of nonzero frequencies. We consider the sub-barrier ionization. The condition to be satisfied is the opposite to the case of pure classical ionization, $F \ll F_B$, in addition we have the quasiclassical condition $\hbar \om \ll E_b$. No restrictions are applied to the parameter $\eps$. Thus we will cover both the regime of relativistic tunnel and multiphoton ionization. 

We employ the relativistic version of the Landau-Dykhne formula \cite{PMK97,DK98}. The ionization probability in quasiclassical approximation and with exponential accuracy reads
\be{prob}
{W} \propto \exp\left\{- \frac{2}{\hbar} \, {\rm Im}~\left(S_f(0;t_0)+S_i(t_0)\right)\right\} \,,
\ee
where $S_i=E_0 t_0$ is the initial part of the action, $S_f$ is given by Eq. (\ref{Srel}). The complex initial time $t_0$ has to be determined from the classical turning point in the complex half-plane \cite{PMK97,DK98}:
\be{cond1}
E_f(t_0)=mc^2\Biggl\{\frac{1+\al^2+f^2}{2 \al} - \frac{e}{m c^2 \al} \DK{f} \DK{A(t_0)}  + \frac{e^2}{2 m^2c^4 \al}  \DK{A}^2(t_0)  \Biggr\}=E_0=mc^2-E_b \,.
\ee
The minimization of the imaginary part of the action leads to the following boundary conditions \cite{PKM67}
\be{cond2}
(x,\DK{r})(t_0)=0\,,~~~
{\rm Im}~(x,\DK{r})(t=0)=0\,.
\ee

In order to obtain simple analytical results we consider the case of {\it linearly} polarized laser light. Then by minimizing the action we obtain from Eqs. (\ref{cond1}) and (\ref{cond2}) that $\DK{f}=0$. Further we obtain a system of nonlinear equations for the determination of complex initial time $t_0$ and constant $\al$,
\beq{t_0}
t_0&=&i \tau_0 = - \frac{i}{\om} \arsinh \left(\eta \sqrt{1+\al^2-2\al \veps_0}\right)\,, \nn\\
\al^2&=&1+\frac{1}{2 \eta^2} \left[1-\frac{\eta \sqrt{1+\al^2-2\al \veps_0}}{\arsinh \left(\eta \sqrt{1+\al^2-2\al \veps_0}\right)}\sqrt{1+\eta^2 \left(1+\al^2-2\al \veps_0 \right)} \right] \,,
\eeq
with the dimensionless initial energy $\veps_o=E_0/mc^2$ and the relativistic adiabatic parameter $\eta = \eps^{-1}= \om mc/eF$. Substituting the values $t_0$ and $\al$ into the final state action we obtain the probability of relativistic quasiclassical ionization in the field of linearly polarized laser light. Within exponential accuracy we get
\be{relprob}
W \propto \exp \Biggl\{-\frac{2 E_b}{\hbar \om} \bigg[ \bigg(1+\frac{1}{2 \gamma^2 \al}+\frac{mc^2}{E_b} \frac{(1-\al)^2}{2\al} \bigg) \arsinh \gamma(\al) -\frac{1}{2 \gamma^2 \al}\gamma(\al) \sqrt{1+\gamma^2(\al)} \bigg] \Biggr\}
\ee
where $\al$ has to be taken as the solution of Eqs. (\ref{t_0}). Further $\gamma = \sqrt{2mE_b}\om/eF$ is the common adiabatic Keldysh parameter from nonrelativistic theory \cite{DK98} and $\gamma(\al)=\eta \sqrt{1+\al^2-2\al \veps_0}$ is an $\al$-depending adiabatic parameter. Equation (\ref{relprob}) is the most general expression for the relativistic ionization rate in the quasiclassical regime and for field strength smaller than the above-barrier threshold. It describes  both the tunnel as well as the multiphoton ionization. It is the relativistic generalization of the famous Keldysh result \cite{Keldysh}. 

Consider now some limiting cases. In the limit of tunnel ionization $\eta \ll 1$ we reproduce the static result of Refs. \cite{PMK97,PMK98} and obtain the first frequency correction
\beq{tunnel}
W &\propto& \exp \bigg\{- \frac{F_{\rm S}}{F} \Phi \bigg\}\,,\nn \\
\Phi&=&\frac{2 \sqrt{3} (1-\al_0^2)^{3/2}}{\al_0} - \frac{3 \sqrt{3} (1-\al_0^2)^{5/2}}{5 \al_0} \eta^2 + O(\eta^4)\,,
\eeq
where $F_s=m^2c^3/e\hbar=1.32 \cdot 10^{16} {\rm V}/{\rm cm}$ is the Schwinger field of quantum electrodynamics \cite{Schwinger} and $\al_0=(\veps_0+\sqrt{\veps_0^2+8})/4$. In the nonrelativistic regime, $\veps_b=E_b/mc^2 \ll 1$, the parameter $\al_0=1-\veps_b/3+\veps_b^2/27$ and the probability of nonrelativistic tunnel ionization including the first relativistic and frequency corrections reads
\beq{tunnelnon}
W \propto \exp \Bigg\{- \frac{4}{3} \frac{\sqrt{2m}E_b^{3/2}}{e\hbar F} \bigg[ 1 -\frac{\gamma^2}{10} - \frac{E_b}{12 mc^2} \bigg(1-\frac{13}{30} \gamma^2 \bigg) \bigg] \Bigg\}\,.
\eeq
Here the first two terms in the brackets describe the familiar nonrelativistic ionization rate including the first frequency correction \cite{DK98}, the next two terms are the first relativistic corrections.  It  follows from Eq.(\ref{tunnel}) that the account of relativistic effects increases the ionization rate in comparison with the nonrelativistic rate. However, even for binding energies of the order of the electron rest energy the relativistic correction in the exponent is quite small. 
 In the ``vacuum'' limit Eq. (\ref{tunnelnon}) results into $W \propto \exp \{- {9 F_{\rm S}}/{2 F}(1-9/40\eta^2)\}$. We find a maximal deviation of about $18\%$ in the argument of the exponential from the Keldysh formula.
 Here the ``vacuum'' limit shall not be confused with the pair creation from the vacuum. It is known that there a no nonlinear vacuum phenomena for a plane wave \cite{Schwinger}. In contrast to that we deal here with the ionization of an atom being in rest in the laboratory system of coordinates. We also mention that we employ the single particle picture. Therefore the pair production processes are beyond the scope of the present paper.

Consider now the multiphoton limit $\eta \gg 1$. In this case the parameter $\al=1-\veps_b/2 \ln 2 \gamma$ and the ionization probability in the relativistic multiphoton limit reads
\be{multi}
W \propto \exp \Biggl\{-\frac{2 E_b}{\hbar \om} \bigg[ \ln 2 \gamma - \frac{1}{2} - \frac{E_b}{8 mc^2 \ln 2 \gamma} \bigg] \Bigg\}\,.
\ee
Again the first two terms in the brackets reflect the nonrelativistic result \cite{Keldysh}, the relativistic effects which lead to an enhancement of the ionization probability are condensed in the third term. 

It has been shown that there is an enhancement of ionization rate in the relativistic theory for both large and small $\eta$.
This should be compared with the results found by Crawford and Reiss.  In their numerical calculations they also found an enhancement of relativistic ionization rate for a circularly polarized field and for $\eta \gg 1$, but for $\eta \ll 1$ their results suggest a strong reduction of the ionization probability \cite{CR94}. For the case of linearly polarized light the ionization rate is found to be reduced by relativistic effects \cite{CR98}. However, Crawford and Reiss studied the above-barrier ionization of hydrogen atom within the strong-field approximation. In contrast to that we have investigated the sub-barrier ionization from a strongly bound electron level, which yields an enhancement of the ionization rate. This enhancement is connected with a smaller initial time $t_0$. As a result the under barrier complex trajectory becomes shorter and the ionization rate increases in comparison with the nonrelativistic theory. Figure 1 shows the relativistic ionization rate Eq. (\ref{relprob})  and the nonrelativistic Keldysh formula as a function of the binding energy $e_b$. The figure should be considered only as an illustration of the enhancement effect. The frequency and intensity parameters used for the calculations are still not available for the experimentalists.

The switch from the multiphoton to the tunnel regime with increasing field strength may be studied in the nonrelativistic limit $\veps_b \ll 1$. Here within first order of $\veps_b$, with $\al=1-(\veps_b/2 \gamma^2)[(\gamma/\arsinh \gamma)\sqrt{1+\gamma^2}-1]$, the ionization probability is found to be
\beq{nonrelprob}
W &\propto& \exp \Bigg\{ - \frac{2 E_b}{\hbar \om} f(\gamma) \Bigg\}\,, \nn \\
f(\gamma)&=& \arsinh \gamma + \frac{1}{2 \gamma^2} \left[ \arsinh \gamma - \gamma \sqrt{1+ \gamma^2} \right] - \veps_b \frac{\gamma^4 + \gamma^2 - 2 \gamma \sqrt{1+\gamma^2} \arsinh \gamma + \arsinh^2 \gamma}{8 \gamma^4 \arsinh \gamma} \,.
\eeq
The terms in $f(\gamma)$ which do not vanish as $\veps_b \to 0$ represent the nonrelativistic quasiclassical ionization rate found by Keldysh \cite{Keldysh}; the terms proportional to $\veps_b$ are the first relativistic correction to the Keldysh formula. Equation (\ref{nonrelprob}) is valid in the whole $\gamma$-domain, i.e., in the multiphoton regime $\gamma < 1$ as well as in the tunnel limit $\gamma > 1$. For small adiabatic parameters, i.e., $\gamma \to 0$, it coincides with Eq. (\ref{tunnelnon}); in the case of large $\gamma \to \infty$ it transforms to Eq.(\ref{multi}). We mention that Eq. (\ref{nonrelprob}) reproduces the full relativistic formula Eq. (\ref{relprob}) with very high accuracy for $E_b < mc^2$.

Consider now the modifications of the energy spectrum induced by relativistic effects. In the nonrelativistic theory and
in the case of linear polarization the most probable value for the electron momentum at the time of emission, $t=0$, is zero. The electrons are preferably emitted in the direction of the polarization of the laser beam. In the relativistic theory employed in this paper we may set the constants $a_1=a_2=0$ in  Eqs. (\ref{trajec}). Then we obtain for the most probable emission velocity in the laboratory system of coordinates
\be{leave}
v_x=c \frac{1-\al^2}{1+\al^2}\,,~~v_y=v_z=0\,,
\ee
where $\al$ has to be taken as the solution of the second equation of Eqs. (\ref{t_0}). In the static limit $\om \to 0$ we reproduce the results of Mur {\it et al.} \cite{PMK98}. It follows from these equations that a strongly bounded electron is emitted in the direction of the laser beam propagation, i.e., perpendicular to the direction of the laser beam polarization. 
 For a nonrelativistic initial state, $\veps_b \ll 1$, the mean emission velocity along the beam propagation $v_x=c\,e_b/3$  is small. Nevertheless, the mean emission velocity seems to be the most sensitive measure of the appearance of relativistic effects in the initial states. In Fig. 2 the $x$-component of the leaving velocity is plotted versus the binding energy of the initial state. Though we have choosen the same parameters of the laser beam as in Fig. 1 it should be mentioned that the dependence of the emission velocity $x$-component on the laser parameters is rather weak. The main parameter determining the leaving velocity along the propagation of the laser beam is the binding energy of the atom.

The electron energy spectrum is also influenced by relativistic final states effects. We put $a_1=p_{y,0}/mc$, $a_2=p_{z,0}/mc$ and $\al=(-p_{x,0}+\sqrt{1+p_{x,0}^2+p_{y,0}^2+p_{z,0}^2})/mc$.  The calculations will be restricted to the tunnel regime $\gamma \ll 1$. Assuming weak relativistic effects in the initial and final states, $\veps_b \ll 1$ and $p_{y,0},~p_{z,0} \ll mc$ one obtains
\be{pprop}
W_p=W \exp\Bigg[- \frac{(p_{x,0}-<p_{x,0}>)^2}{m} \frac{\gamma}{\hbar \om} - \frac{p_{z,0}^2}{m} \frac{\gamma}{\hbar \om} \Bigg]\, \exp \Bigg[- \frac{p_{y,0}^2}{3m} \frac{\gamma^3}{\hbar \om}-  \frac{p_{y,0}^4}{4m^3c^2} \frac{\gamma}{\hbar \om} \Bigg]\,,
\ee
where $W$ is the total ionization rate Eq. (\ref{tunnelnon}) in the weak relativistic tunnel regime. The first exponent in Eq.(\ref{pprop}) describes  the momentum distribution in the plane perpendicular to the polarization axis. There is only one relativistic effect in the weak relativistic regime considered here - the appearance of the  mean momentum at the emission time $<p_{x,0}>=E_b/3c$. The nonzero mean emission velocity along the propagation vector destroys the symmetry in the (x,z)-plane that exists in non-relativistic theory. The first term in the second exponent of Eq.(\ref{pprop}) determines the nonrelativistic energy spectrum for the low energetic electrons moving along the polarization axis $ p_{y,0}^2 < 4 \gamma^2 m^2 c^2/3$, whereas the second, relativistic term  becomes important for the high energy tail $ p_{y,0}^2 > 4 \gamma^2 m^2 c^2/3$. It is only in the case of small adiabatic parameter $\gamma \le 0.1$, that the high energy condition does not contradict the condition $ p_{y,0}<m c$.  We mention that the second term in the second exponent agrees with a corresponding term of Krainov \cite{K98,DK98}.

  In conclusion, 
the expressions obtained in this paper within exponential accuracy may be improved by taking into account the Coulomb interaction through the perturbation theory. The results of this paper may be also used in nuclear physics and quantum chromodynamics.

This research was partially supported by the Deutsche Forschungsgemeinschaft (Germany).

\newpage

\begin{center}
{\bf FIGURE CAPTIONS}
\end{center}

\begin{description}

\item[(Figure 1)] Absolute value of the logarithm of the ionization rate $-\ln\; W$ versus the binding energy of initial level $e_b=E_b/mc^2$. The solid line shows the relativistic rate Eq.(\ref{relprob}), the dashed line is the nonrelativistic Keldysh formula (Eq. (\ref{nonrelprob}) without the relativistic correction term). The curves are shown for a frequency $\om = 100$ and an intensity $I=8.5 \cdot 10^{7}$ (in a.u.).
\item[(Figure 2)] The $x$-component of the emission velocity {$v_x/c$} versus the binding energy of initial level {$e_b=E_b/mc^2$}.  The emission velocity in the nonrelativistic theory is zero. The curve is shown for a frequency $\om = 100$ and an intensity $I=8.5 \cdot 10^{7}$ (in a.u.).

\end{description}


\begin {figure} [h] 
\unitlength1mm
  \begin{picture}(155,160)
\put (0,10){\psfig{figure=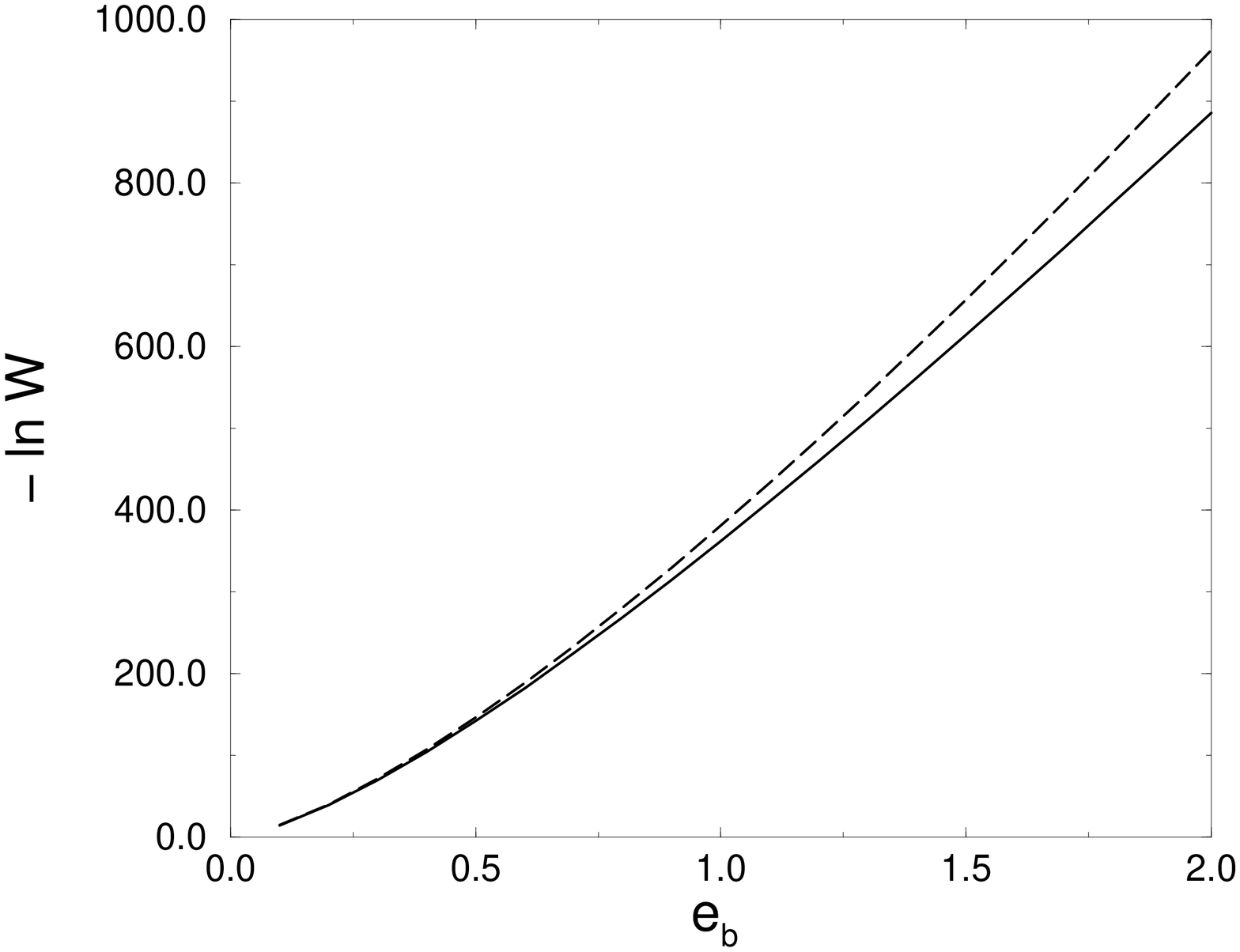,width=15.0cm,height=13.0cm,angle=-90}}
 \end{picture}\par
\caption{}
\end{figure}

\begin {figure} [h] 
\unitlength1mm
  \begin{picture}(155,160)
\put (0,10){\psfig{figure=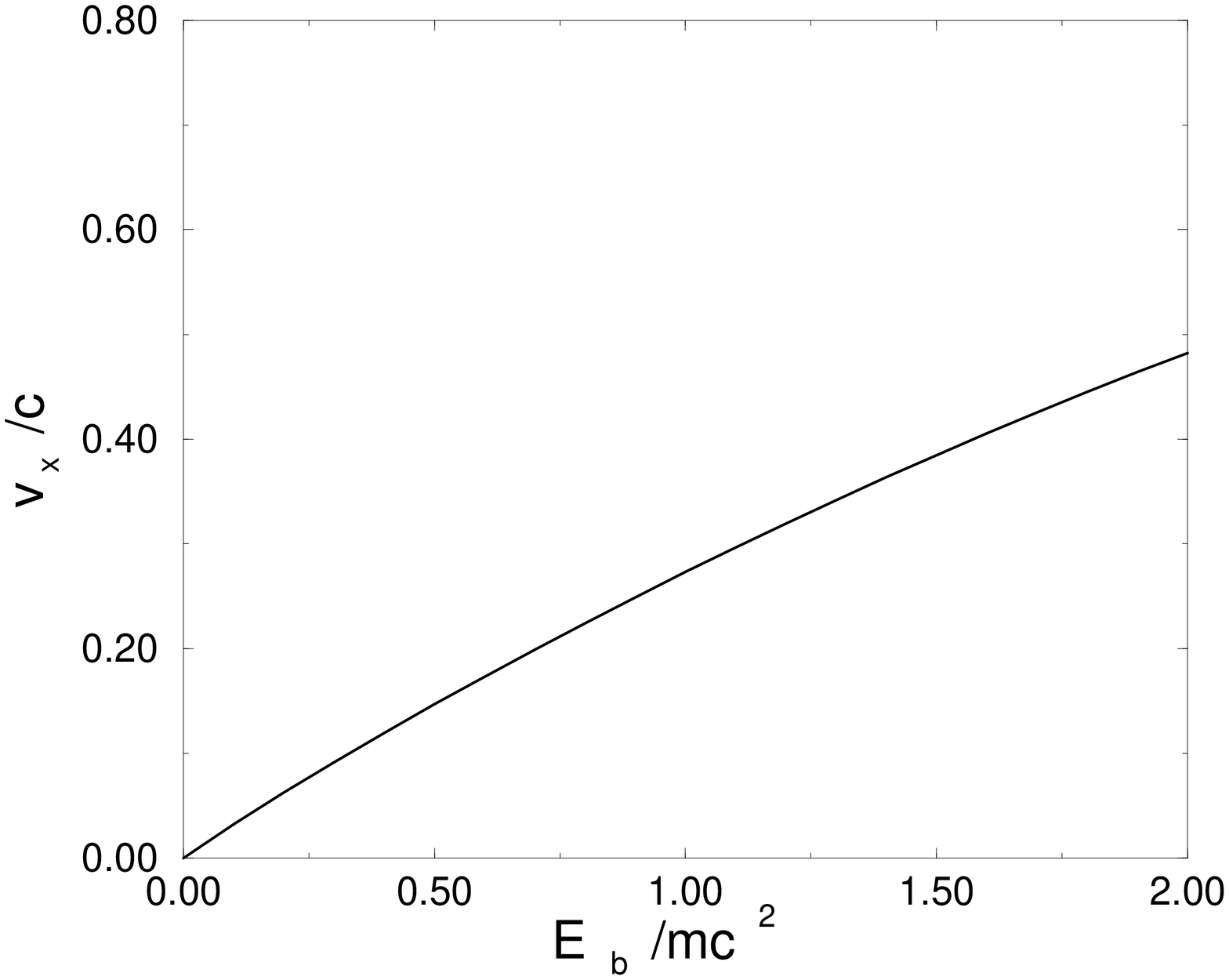,width=15.0cm,height=13.0cm,angle=-90}}
 \end{picture}\par
\caption{}
\end{figure}

\end{document}